\documentclass[reprint,amsmath,longbibliography,amssymb,aps]{revtex4-1}

\usepackage{MnSymbol}
\usepackage{graphicx}
\usepackage{subfigure}
\usepackage{epstopdf}
\usepackage{hyperref}
\usepackage{bm}
\usepackage{color}
\usepackage[disable]{todonotes}
\usepackage{braket}
\usepackage{xcolor}
\usepackage{gensymb}
\usepackage{siunitx}
\usepackage{placeins}
\usepackage[compat=1.0.0]{tikz-feynman}

\begin{document}

\title{Phonon-mediated superconductivity in doped monolayer materials }

\author{Even Thingstad}
\author{Akashdeep Kamra}
\author{Justin W. Wells}
\author{Asle Sudb\o}
\affiliation{Center for Quantum Spintronics, Department of Physics, Norwegian University of Science and Technology, NO-7491 Trondheim, Norway}

\date{\today}

\begin{abstract}
Insight into why superconductivity in pristine and doped monolayer graphene seems strongly suppressed has been central for the recent years' various creative approaches to realize superconductivity in graphene and graphene-like systems. We provide further insight by studying electron-phonon coupling and superconductivity in doped monolayer graphene and hexagonal boron nitride based on intrinsic phonon modes. Solving the graphene gap equation using a detailed model for the effective attraction based on electron tight binding and phonon force constant models, the various system parameters can be tuned at will. Consistent with results in the literature, we find slight gap modulations along the Fermi surface, and the high energy phonon modes are shown to be the most significant for the superconductivity instability. The Coulomb interaction plays a major role in suppressing superconductivity at realistic dopings.  Motivated by the direct onset of a large density of states at the Fermi surface for small charge dopings in hexagonal boron nitride, we also calculate the dimensionless electron-phonon coupling strength there, but the comparatively large density of states cannot immediately be capitalized on, and the charge doping necessary to obtain significant electron-phonon coupling is similar to the value in graphene.  
\end{abstract}

\maketitle

\section{Introduction}

The discovery of graphene has attracted massive attention in condensed matter physics, stimulating an enormous number of theoretical and experimental investigations into a class of novel materials broadly denoted as Dirac materials~\cite{Novoselov2004, Novoselov2005, Geim2007, CastroNeto2009, DasSarma2011}. 
Among their interesting properties is the Dirac-cone shaped electron band structure at half filling, enabling the study of relativistic physics in a condensed matter setting~\cite{Geim2007, Katsnelson2006a, Katsnelson2006b, Kim2017}. However, the cone structure with a vanishing density of states and Fermi-surface at the Dirac point suppresses phenomena such as superconductivity, which precisely rely on the existence of a Fermi-surface. 

In spite of this obstacle, there is a plethora of graphene-like systems where superconductivity has been predicted or observed. In carbon nanotubes and the carbon based fullerene crystals also known as ``buckyballs'', superconductivity was observed already decades ago in crystals intercalated with potassium~\cite{Kelty1991, Ganin2010}.  Superconductivity is also well known in graphite intercalation compounds \cite{Hannay1965, Calandra2005, Csanyi2005, Weller2005, Emery2005}, where the interlayer interactions and the additional dopant phonon modes enhance superconductivity~\cite{Calandra2005}. A similar picture arises also for intercalated bilayer graphene, where interlayer interactions are crucial for the resulting superconductivity \cite{Mazin2010, Margine2016}. In bilayer graphene, a different route to superconductivity  is the magic angle twist approach~\cite{Morell2010, Bistritzer2011, Cao2018, Wu2018}, where strong correlations are believed to play a key role. Superconductivity has also been demonstrated in effectively one-dimensional carbon nanotubes~\cite{Kociak2001, Tang2001}, which have strong screening of the repulsive Coulomb interaction. In addition to these intrinsic mechanisms, superconductivity may also be induced by proximity~\cite{Heersche2007, Shailos2007, Du2008, Lee2018}. There, the resulting superconductivity in graphene will necessarily inherit extrinsic key properties from the superconductor it is placed in proximity to \cite{Lee2018}. 

Although superconductivity is already well established in a multitude of these graphene-like systems, its observation in monolayer graphene has proven very challenging. For phonon-mediated superconductivity, the key quantity is the dimensionless electron-phonon coupling (EPC) strength \(\lambda\), which is determined by both the density of states at the Fermi level and the strength of the effective phonon-mediated potential. The first challenge that has to be overcome is therefore doping the system away from the Dirac point. The primary ways of doing this are chemical doping~\cite{Panchakarla2009, Rao2009, Usachov2011, Liu2011, Agnoli2016} and deposition of elements onto (or under) the graphene sheet  \cite{Calandra2007, Bianchi2010, McChesney2010, Profeta2012, Vinogradov2012, Pletikosic2012, Vinogradov2013,  Haberer2013, Si2013, Hwang2014, Ludbrook2015}. Using these methods, doping levels approaching the van Hove singularity have been achieved \cite{McChesney2010}. Second, one must make sure that \(\lambda\) has a sufficiently large value at the achievable doping. Additional dopant phonon modes and modifications of the electron band structure in decorated monolayer graphene may enhance the electron-phonon coupling strength \cite{Profeta2012, Si2013, Hwang2014, Margine2014, Ludbrook2015}, and in these systems, one has even observed some evidence~\cite{Ludbrook2015, Chapman2016} for the desired monolayer graphene superconductivity.

The EPC strength \(\lambda\) can be measured by examining kinks and broadening in the electronic band structure using angular resolved photo-emission spectroscopy (ARPES)~\cite{Grimvall1981, Hellsing2002, Mazzola2017}. At realistic doping energies in the vicinity of the van Hove singularity in the \(\pi\)-band, $\lambda$-values of the same order as in many known conventional superconductors~\cite{Meservey1969} have been predicted and measured experimentally~\cite{Park2008, Johannsen2013, Mazzola2017}. In light of this, superconductivity in single-layer graphene with reasonable critical temperatures does not seem inconceivable even without dopant phonon modes and special electron band structure modifications. Why superconductivity in monolayer graphene remains so hard to achieve is therefore not entirely clear.

In conventional superconducting materials, the Coulomb interaction does not play a significant  role in reducing $T_c$, since the effective phonon-mediated potential is attractive only in a small region around the Fermi surface, whereas the repulsive Coulomb interaction has much longer Brillouin zone variations. The mechanism at work, retardation, can be seen by solving the gap equation with a simplified  model~\cite{Morel1962} for the combined potential. In graphene, however, we do expect the Coulomb interaction~\cite{Kotov2012} to reduce the critical temperature significantly~\cite{Wehling2011, Einekel2011, Si2013, Margine2014} due to the modest electron-phonon coupling strength. A crude estimate of the Morel-Anderson renormalization of the dimensionless EPC strength \(\lambda\) shows that the renormalization is of the same order as \(\lambda\) itself. A detailed study of phonon-mediated superconductivity in graphene is therefore necessary. 

Eliashberg theory for doped monolayer graphene was developed in Ref.~\onlinecite{Einekel2011}, where the pair scattering processes within and between the Fermi surface segments centered around the inequivalent Brillouin zone points \(K\) and \(K'\) were accounted for explicitly. The authors estimate the critical temperature based on the assumption of an isotropic gap, Fermi surface averaged Coulomb interaction \cite{Kotov2012, Si2013} in the linear spectrum regime, and various estimates for the electron-phonon coupling strength based on other works. The resulting critical temperature is of order \(\SI{10}{\kelvin}\) with the optimistic estimates.

In Ref.~\onlinecite{Margine2014}, the electron-phonon coupling strength and superconducting gap were calculated with anisotropic Eliashberg theory based on \textit{ab initio} calculation of the quasimomentum and energy dependent electron-phonon coupling strength. Coulomb interaction effects are incorporated through a Morel-Anderson pseudopotential, which is treated as a semi-empirical parameter. For \(n\)-type doping, they find that superconductivity may be possible due to the presence of a free-electron-like (FEL) band. For \(p\)-type doping, this band is not present, and the Coulomb interaction seems likely to suppress superconductivity.

In this paper, we perform detailed calculations of the electron-phonon coupling based on an electron tight binding model and a phonon force constant model in the presence of a Hubbard-type Coulomb interaction. We then solve an anisotropic BCS-type gap equation, which should give reasonable estimates for the superconducting properties due to the relatively modest electron-phonon coupling strength. The various system parameters in our model can easily be tuned to investigate how various physical mechanisms affect the superconducting properties. Understanding this is essential in the pursuit of realizing superconductivity in monolayer graphene based on the intrinsic in-plane phonon modes.

Our results show that superconductivity with an experimentally measurable gap may be possible for large dopings approaching the van Hove singularity. We find an electron phonon coupling strength and gap anisotropy qualitatively similar and of the same order as in Ref.~\onlinecite{Margine2014}, and the Coulomb interaction is shown to be crucial in reducing the critical temperature of the system. We also look into the contributions to the electron-phonon coupling from the various phonon modes in the system~\cite{Si2013}, and identify the high-energy phonons as the most significant for the superconducting instability in the realistically achievable doping regime.

The two-dimensional material hexagonal boron nitride (h-BN) was discovered shortly \cite{Novoselov2005} after graphene \cite{Novoselov2004} using the same micromechanical cleavage technique to exfoliate monolayers from the stacks of weakly interacting layers also known as van der Waals materials. In many respects, the two are very similar~\cite{Mishra2018}. They have the same lattice structure and a similar lattice constant, which makes h-BN a good substrate for graphene \cite{Dean2010, Wang2017} and suitable for graphene heterostructure engineering~\cite{Yankowitz2019}. Like graphene, it also has strong chemical bonding, and a comparable phonon Debye frequency~\cite{Wirtz2003}. Unlike graphene, however, boron nitride has two different ions, boron and nitrogen, on the two honeycomb sublattices. This has dramatic consequences for the electronic band structure, since the Dirac cone in graphene is protected by time reversal and inversion symmetry. Breaking of the latter symmetry therefore renders hexagonal boron nitride a large gap insulator~\cite{Robertson1984}.

The possibility of superconductivity in doped hexagonal boron nitride is a lot less studied than in doped graphene, but a recent density functional theory study~\cite{Shimada2017} suggests that decorated h-BN may become superconducting with a transition temperature of up to \(\SI{25}{\kelvin}\). Although the dopant phonon modes are again responsible for this relatively large transition temperature, this also hints at possibilities for superconductivity mediated by intrinsic in-plane phonon modes. Furthermore, and very different from graphene, the parabolic nature of the electron band close to the valence band maximum gives a direct onset of a large density of states even at small charge doping. Motivated by this, we use the same methodology as in the graphene case to calculate the dimensionless electron-phonon coupling strength \(\lambda\) for hexagonal boron nitride. Due to suppression of the electron-phonon coupling matrix element due to the small Fermi surface, however, this effect cannot be capitalized on, and we find that h-BN has an electron-phonon coupling strength similar to graphene.

In Sec. \ref{sec_electronsPhononsElPhCoupling} of this paper, we first present the free electron and the free phonon models for graphene briefly, followed by a more thorough derivation of the tight binding electron-phonon coupling. In Sec. \ref{sec_coulomb}, we introduce and discuss the Hubbard-type Coulomb interaction used in this paper. In Sec. \ref{sec_gapEq}, we introduce the assumed pairing, resulting gap equation and effective phonon-mediated potential, before presenting the numerical results for graphene in Sec. \ref{sec_numericalResults}.
In Sec. \ref{sec_discussion}, we discuss some qualitative aspects of these results. Switching to boron nitride in Sec. \ref{sec_boronNitride}, we discuss how the opening of a gap changes the band structure and electron-phonon coupling. Finally, the paper is summarized in Sec. \ref{sec_summary}.

\section{Electrons, phonons, and electron-phonon coupling}\label{sec_electronsPhononsElPhCoupling}

We consider a model for electrons on the graphene lattice, and allow for lattice site vibrations.
For the electrons, we use a nearest neighbour tight binding model~\cite{CastroNeto2009} describing the \(\pi\)-bands, as explained in further detail in Appendix \ref{sec_appendix_electron}. Other bands are disregarded, since only the \(\pi\)-bands are close to the Fermi surface for realistically achievable doping levels in graphene. For the phonons, we use a force constant model with nearest and next-to-nearest neighbour couplings as introduced in Refs.~\onlinecite{Falkovsky2007, Falkovsky2008} and elaborated in Appendices \ref{sec_appendix_phonon} and \ref{sec_appendix_forceConstants}. These models give a realistic band structure and realistic phonon spectra.

The electron-phonon coupling model is derived by assuming the electrons to follow the lattice site ions adiabatically, and by Taylor expanding the overlap integral \(t_{ij} \) in the hopping Hamiltonian

\begin{equation}\label{eq_tightBinding}
H = - \sum_{\langle i, j \rangle, \sigma} ( t_{ij} c_{i\sigma}^\dagger c_{j\sigma} + \mathrm{h.c.} )
\end{equation}

\noindent to linear order in the deviations. Here, \(c_{i\sigma}^\dagger\) and \(c_{i\sigma}\) are creation and annihilation operators for an electron at site \(i\) with spin \(\sigma \in \{\uparrow, \downarrow\}\). Considering only the nearest neighbour hoppings, we obtain

\begin{equation}
t_{i+\pmb{\delta}_A, i} = t_1 + (\mathbf{u}_{i+\pmb{\delta}_A} - \mathbf{u}_{i}) \cdot \nabla_{\pmb{\delta}} t_1(\pmb{\delta}),
\end{equation}

\noindent where \(t_1\) is the nearest neighbour hopping amplitude, \(\mathbf{u}_i\) is the ionic displacement of lattice site \(i\) from its equilibrium position, and the overlap integral \(t_1(\pmb{\delta})\) is regarded as a function of the relative position \(\pmb{\delta}\) of the two lattice sites \(i\) and \(i+\pmb{\delta}_A\), where \(\pmb{\delta}_A\) is the equilibrium nearest neighbour vector from the \(A\) to the \(B\) sublattice. Due to the mirror-symmetry about the line connecting the lattice sites \(i\) and \(i+\pmb{\delta}_A\), the electron-phonon coupling can be written as

\begin{equation}
H_\mathrm{el-ph} =   \frac{\gamma t_1}{d^2} 
\sum_{i\in A, \pmb{\delta}_A, \sigma}  
\pmb{\delta}_A \cdot  (\mathbf{u}_{i+\pmb{\delta}_A} -\mathbf{u}_i ) (c_{i+\pmb{\delta}_A,\sigma}^\dagger c_{i,\sigma} + \mathrm{h.c.} ),
\end{equation}

\noindent where \(\gamma = - {\mathrm{d} \ln t_1 }/{\mathrm{d} \ln d}\) is a dimensionless number of order 1, and $d$ is the equilibrium nearest neighbour distance, which we use as our unit of length. In quasimomentum-space, this gives the electron-phonon coupling


\begin{align}
H_\mathrm{el-ph} &= 
\sum_{\mathbf{k},\mathbf{q}} \sum_{\eta\eta'} \sum_{\nu,\sigma} 
g_{\mathbf{k},\mathbf{k}+\mathbf{q}}^{\eta\eta',\nu}
(a_{\mathbf{q}\nu} + a_{-\mathbf{q},\nu}^\dagger ) c_{\eta'\sigma}^\dagger(\mathbf{k}+\mathbf{q}) 
c_{\eta\sigma}(\mathbf{k}) ,
\end{align}

\noindent where \(a_{-\mathbf{q},\nu}^\dagger\) and \(a_{\mathbf{q}\nu}\) are creation and annihilation operators for in-plane phonons labelled by \(\nu \in \{0,1,2,3\}\), and \(\eta, \eta' = \pm\) denote electron bands.  To linear order in the lattice site deviations, the out-of-plane phonon modes do not couple to the electrons due to the assumed \(z\rightarrow -z\) mirror symmetry of the system \footnote{Since we consider superconductivity in graphene with Fermi surface doped down towards the van Hove singularity in the \(\pi\)-band, only the phonon scattering processes where an electron is scattered within the \(\pi\)-band are of interest. The effect of the mirror symmetry \(z \rightarrow - z\) on the sign of the electron-phonon coupling element \(g\) is determined by the product of the deviation and the incoming and outgoing electron states. Since all these are anti-symmetric under \(z \rightarrow -z\), the out-of-plane phonon modes cannot cause electron transitions within the \(\pi\)-band.}. The coupling matrix element \(g^{\eta\eta',\nu}_{\mathbf{k},\mathbf{k}+\mathbf{q}}\) is given by


\begin{align}\label{eq_elPhCouplingMatrixElement}
g_{\mathbf{k},\mathbf{k}+\mathbf{q}}^{\eta\eta',\nu} = &
\frac{g_0}{\sqrt{N_A}} \sqrt{ \frac{\omega_\Gamma}{\omega_{\mathbf{q}\nu}} } \sum_{\pmb{\delta}_A} \left( \frac{\pmb{\delta}_A}{d} \right) 
 \left[ e^{i\mathbf{q} \cdot \pmb{\delta}_A} \mathbf{e}_\nu^B(\mathbf{q}) - \mathbf{e}_\nu^A(\mathbf{q}) \right] \nonumber \\
 &\times \Big[ e^{i \mathbf{k}\cdot \pmb{\delta}_A } F_{A\eta'}^*(\mathbf{k}+\mathbf{q}) F_{B\eta}(\mathbf{k} ) \nonumber
  \\ & +  e^{-i (\mathbf{k}+\mathbf{q})\cdot \pmb{\delta}_A } F_{B\eta'}^*(\mathbf{k}+\mathbf{q}) F_{A\eta}(\mathbf{k} ) \Big],
\end{align}

\noindent  where \(F_{D\eta}(\mathbf{k})\) is the sublattice amplitude of electron band \(\eta\) at quasimomentum \(\mathbf{k}\) and follows from the diagonalization of the free electron model, as elaborated in Appendix \ref{sec_appendix_electron}. Similarly, \(\mathbf{e}_\nu^D(\mathbf{q})\) is the phonon polarization vector at sublattice \(D\in \{A,B\}\) for the phonon mode \((\mathbf{q},\nu)\), and follows from diagonalization of the in-plane phonon Hamiltonian (see Appendix \ref{sec_appendix_phonon} for details). The phonon mode frequencies are denoted by  \(\omega_{\mathbf{q}\nu}\), \(N_A\) is the number of lattice sites on the A sublattice, and the energy scale \(g_0\) is given by

\begin{equation}
g_0=\sqrt{\left(\frac{\hbar^2}{2Md^2} \right) \frac{1}{ \hbar\omega_\Gamma}}  \gamma t_1,
\end{equation}

\noindent where \(M\) is the carbon atom mass, and \(\omega_\Gamma\) is a phonon energy scale given by the optical phonon frequency at the \(\Gamma\)-point \(\mathbf{q} = (0,0)\).

To quantify the strength of the electron-phonon coupling, one may introduce the dimensionless electron-phonon coupling strength parameter~\cite{Mahan2000, Hellsing2002}

\begin{equation}
\lambda_{\mathbf{k}\eta} = \sum_{\mathbf{q} \nu}  \frac{2}{\hbar \omega_{\mathbf{q}\nu}} |g^{\eta \eta,\nu}_{\mathbf{k},\mathbf{k}+\mathbf{q}}|^2 \delta(\epsilon_{\mathbf{k}+\mathbf{q}, \eta}-\epsilon_{\mathbf{k}\eta}),
\end{equation}

\noindent where \(\epsilon_{\mathbf{k}\eta}\) is the electron single particle energy. We have neglected interband scattering processes since the \(\pi\)-band only overlaps with the lower lying \(\sigma\)-bands at unrealistic doping levels~\cite{Hellsing2018, Mazzola2017}.

Averaging \(\lambda_{\mathbf{k}\eta}\) over the Fermi surface corresponding to the energy of the incoming momentum often provides a simple and useful tool for understanding the dependence of the critical temperature of a superconductor on other system parameters through the BCS formula
\(k_B T_c \approx \hbar \omega_D \exp (-1/\lambda)\), where \(\lambda\) is the Fermi surface average of \(\lambda_{\mathbf{k}\eta}\).

\section{Coulomb Interaction}\label{sec_coulomb}

To include the effect of the Coulomb interaction, we use the repulsive Hubbard interaction

\begin{equation}\label{eq_coulombRealSpace}
V^C = u_{0} \sum_i n_{i\uparrow} n_{i\downarrow},
\end{equation}

\noindent where \(n_{i\sigma}\) is the electron number operator. The on-site repulsion \(u_0\) has been calculated from \textit{ab initio} in undoped  graphene~\cite{Wehling2011}. At significant doping of order \(\SI{2}{\eV}\), as discussed in Appendix \ref{sec_appendix_coulomb}, the screening length is a small fraction of the nearest neighbour bond length, and we therefore disregard longer ranged interactions. 

For doped graphene, we expect the onset of \(\pi\)-band screening to reduce the on-site repulsion. A simple model for \(u_0(\mu)\) is obtained by calculating the polarization bubble in the linear spectrum approximation for intra-valley scattering processes~\cite{Kotov2012}. The resulting polarization bubble is momentum independent, and this gives

\begin{equation}\label{eq_onsiteHubbard}
u_0(\mu) = \frac{u_0(0)}{1 + \alpha u_0(0) \rho(\mu) A_\mathrm{cell}},
\end{equation}

\noindent where \(\rho(\mu)\) is the density of states per area in the linear spectrum approximation, and \(A_\mathrm{cell}\) is the real space area associated with the unit cell. We have introduced a factor \(\alpha\) to be able to study polarization strength dependence. The doping dependence can also be interpreted as an interpolation between the known on-site Coulomb repulsion \(u_0(0)\) for pristine graphene, and the known result of doping independent Coulomb pseudo-potential \(\mu_C\) at the Fermi surface~\cite{Si2013, Margine2014}, requiring \(u_0 \propto 1/\rho(\mu)\).


In momentum space, the Coulomb interaction takes the form

\begin{equation}\label{eq_coulombStructureFactorsBCS}
V^C = 
\frac{u_0}{2N_A} 
\sum_{\mathbf{k}\tilde{\mathbf{k}} \mathbf{q}}
\sum_{\eta_1 \cdots \eta_4}
c_{\eta_1 \uparrow}^\dagger({\mathbf{k}} + \mathbf{q}) 
c_{\eta_2 \downarrow}^\dagger(\tilde{\mathbf{k}} - \mathbf{q}) 
c_{\eta_3 \downarrow}(\tilde{\mathbf{k}})
c_{\eta_4 \uparrow}(\mathbf{k})
\end{equation}

\noindent in terms of the momentum band basis annihilation operators \(c_{\eta\sigma}(\mathbf{k})\).

\section{Pairing and gap equation}\label{sec_gapEq}

\noindent 
The in-plane phonons yield an 
effective interaction between the electrons in the system that may cause pairing and superconductivity. Assuming spin-singlet pairing at \(\pm \mathbf{k}\) and considering only the electron band \(\pi^-\), the relevant interaction can be written in the form

\begin{equation}
V = \sum_{\mathbf{k} \mathbf{k}'} V_{\mathbf{k}\mathbf{k}'} 
c_{-\uparrow}^\dagger(\mathbf{k}') c_{-\downarrow}^\dagger(-\mathbf{k}') c_{-\downarrow}(-\mathbf{k}) c_{-\uparrow}(\mathbf{k})
\end{equation}

\noindent with a potential \(V_{\mathbf{k}\mathbf{k}'}\) that contains contributions both from the Coulomb potential and an effective phonon-mediated potential \(V_{\mathbf{k}\mathbf{k}'}^\mathrm{ph-m.}\), so that

\begin{equation}
V_{\mathbf{k}\mathbf{k}'} = V_{\mathbf{k}\mathbf{k}'}^{C} + V_{\mathbf{k}\mathbf{k}'}^\mathrm{ph-m.}.
\end{equation}

The Coulomb contribution is given by Eq. \eqref{eq_coulombStructureFactorsBCS}. The effective phonon-mediated potential follows from a canonical transformation \cite{Heid2017}, and is given by

\begin{equation}
V_{\mathbf{k}\mathbf{k}'}^\mathrm{ph-m.} =  \sum_\nu |g_{\mathbf{k},\mathbf{k}+\mathbf{q}}^{--,\nu}|^2 
\frac{2 \hbar \omega_{\mathbf{q}\nu}}
{(\epsilon_{\mathbf{k} + \mathbf{q}} - \epsilon_\mathbf{k} )^2  - (\hbar\omega_\mathbf{q\nu})^2},
\end{equation}

\noindent where the quasimomentum \(\mathbf{q}\) is defined by \(\mathbf{k}'=\mathbf{k}+\mathbf{q}\).

Due to the singlet pairing assumption, the gap has to be symmetric under \(\mathbf{k} \rightarrow - \mathbf{k}\), and therefore, the potential \(V_{\mathbf{k}\mathbf{k}'}\) can be replaced with the symmetrized potential

\begin{equation}
V_{\mathbf{k},\mathbf{k}'}^\mathrm{symm} = \frac{1}{2} \left( V_{\mathbf{k},\mathbf{k'}} + V_{\mathbf{k},-\mathbf{k}'} \right),   
\end{equation}

\noindent which is symmetric under \(\mathbf{k}\rightarrow -\mathbf{k}\) and \(\mathbf{k}' \rightarrow -\mathbf{k}'\), as well as interchange of the incoming and outgoing momenta \(\mathbf{k}\) and \(\mathbf{k}'\).

To proceed, we have to solve the gap equation

\begin{equation}
\Delta_\mathbf{k} =  - \sum_{\mathbf{k}'} V_{\mathbf{k}\mathbf{k}'}^\mathrm{symm} \chi_{\mathbf{k}'} \Delta_{\mathbf{k}'} ,
\label{eq_gapEquation}
\end{equation}

\noindent with susceptibility 

\begin{equation}
\chi_{\mathbf{k}} = \frac{\tanh \beta E_\mathbf{k} /2}{2E_\mathbf{k}} , \qquad E_\mathbf{k} = \sqrt{ \xi_\mathbf{k}^2 + |\Delta_\mathbf{k}|^2 },
\end{equation}

\noindent where \(E_\mathbf{k}\) is the quasiparticle excitation energy, and \( \xi_\mathbf{k}=\epsilon_\mathbf{k} - \mu\) is the single particle energy \(\epsilon_\mathbf{k}\) measured relative to the Fermi surface at chemical potential \(\mu\).

To find the critical temperature and the gap structure \(\Delta_\mathbf{k}\) just below the critical temperature, it suffices to neglect the gap in the excitation spectrum \(E_\mathbf{k}\) in the gap equation. This gives an eigenvalue problem linear in the eigenvectors and non-linear in the eigenvalue, which is solved as discussed in Appendix \ref{sec_appendix_numericalDetails} to obtain the critical temperature and gap momentum dependence.

\begin{figure*}[htb]
    \centering
    \includegraphics[width=0.9\textwidth]{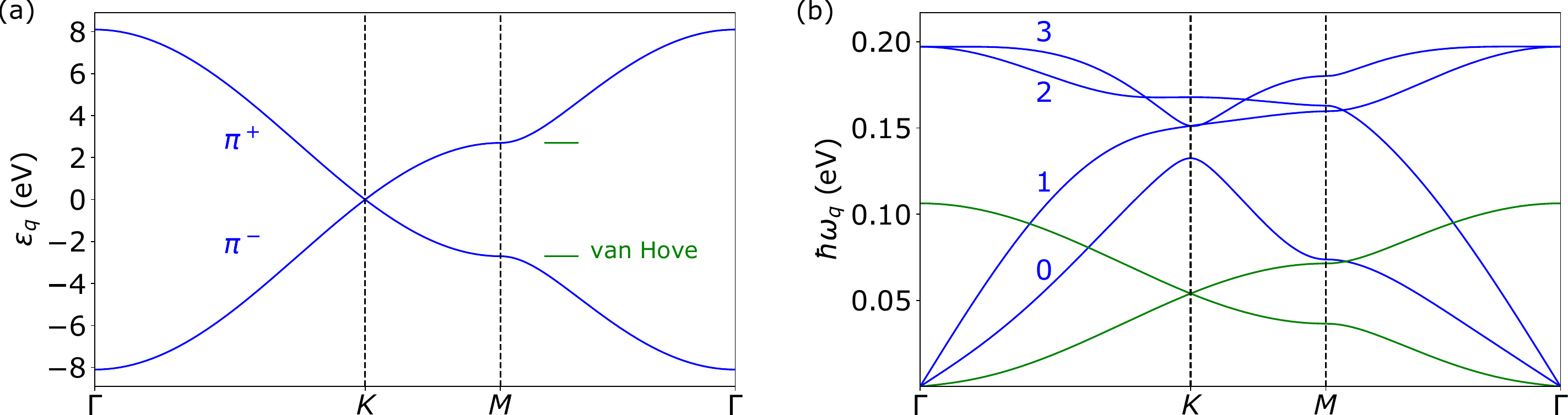}
    \caption{(a) Electron spectrum for the \(\pi\)-bands of graphene in a tight binding hopping model. (b) Phonon spectrum for free-standing graphene in the force constant model. In-plane modes are shown in blue, with out-of-plane modes in green. At any point in the Brillouin zone, the in-plane phonon modes are labelled according to energy. }
    \label{fig_freeSpectra}
\end{figure*}

\section{Graphene numerical results}\label{sec_numericalResults}

\subsection{Parameter values and free spectra}

We set the equilibrium electron hopping amplitude \(t_1\) to \(2.8\; \mathrm{eV}\) \cite{Kotov2012}. The resulting electron band structure for the \(\pi\)-bands of graphene is shown in Fig \ref{fig_freeSpectra}(a).  For the phonon force constant model used to derive the phonon spectrum, we use the same parameter values as Ref.~\onlinecite{Falkovsky2007}, and the resulting excitation spectrum is shown in Fig.~\ref{fig_freeSpectra}(b).

The dimensionless parameter \(\gamma\) can be estimated from \textit{ab initio}, and is roughly \(2.5\) \cite{Frederiksen2020}. This gives reasonable values~\cite{Park2008} for the dimensionless electron-phonon coupling strength \(\lambda\). With phonon energy scale \(\hbar \omega_\Gamma = 0.20 \; \mathrm{eV}\) and nearest neighbour distance \(d=1.42\; \mathrm{\AA}\)~\cite{Katsnelson2012}, this gives \(g_0=0.15 \; \mathrm{eV}\). All system parameters involved in the calculation of the energy scale are tabulated in Appendix \ref{sec_appendixParameterValues}.

\subsection{Electron-phonon coupling strength and effective potential}

\begin{figure*}
    \centering
    \includegraphics[width=1\textwidth]{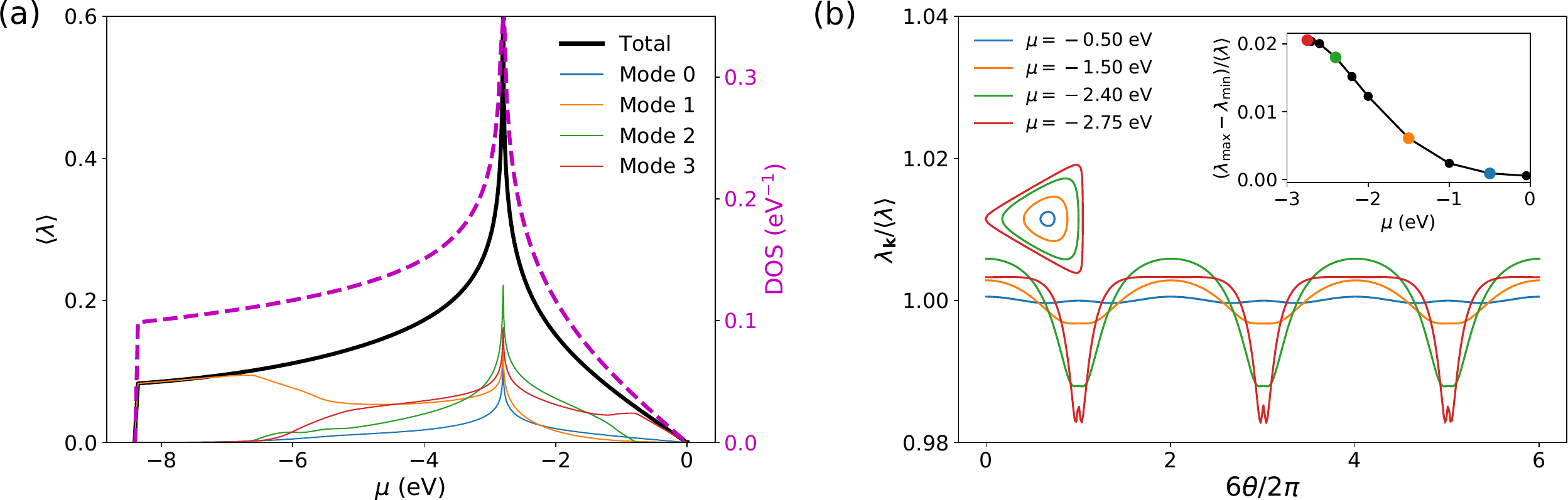}
    
    \caption{(a) Calculated electron-phonon coupling strength \(\lambda\) averaged over the Fermi surface (black) and electronic density of states (magenta) as function of chemical potential. Since \(\lambda\) is highly dependent on the density of states but also dependent on the electron-phonon coupling element \(|g_{\mathbf{k}\mathbf{k}'}^{--,\nu}|\), the two have similar but not identical shapes. The contributions to \(\lambda\) from the various in-plane phonon modes are shown in colors. (b) Electron phonon coupling strength along the Fermi surface normalized to the mean value for various doping levels. As shown in the inset, the electron-phonon coupling strength modulations are increasing with doping toward the van Hove singularity. }
    \label{fig_lambda}
\end{figure*}

\begin{figure*}
\centering
\includegraphics[width=0.95\textwidth]{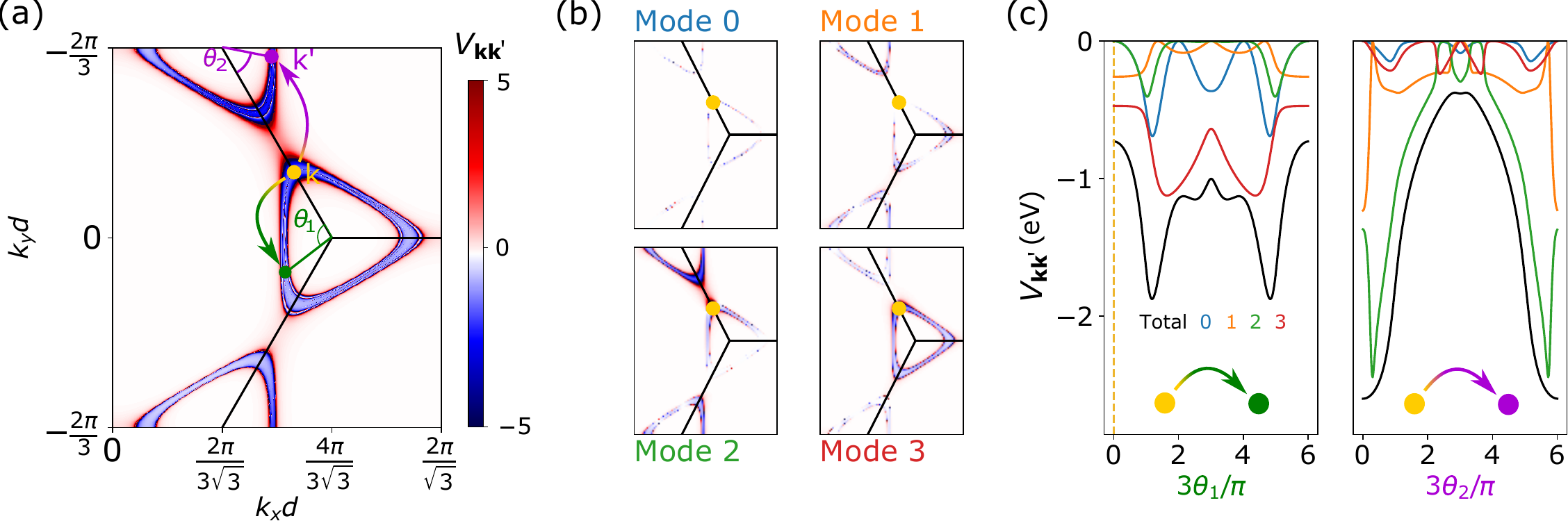}
\caption{ (a) Graphene effective interaction \(V_{\mathbf{k}\mathbf{k}'}^\mathrm{ph-m.}\) in eV for incoming electron momentum \(\mathbf{k}\) (yellow dot) at outgoing momentum \(\mathbf{k}'\) indicated by position in the plot. There can be both intra- (green) and inter-valley (magenta) scattering processes to the Fermi surface. The potential is attractive close to the Fermi surface, before it turns repulsive at a characteristic phonon frequency, and then decays to zero. (b) Decomposition of effective potential in phonon mode contributions. The size of the attractive region clearly depends on the phonon energy. (c) Effective potential \(V_{\mathbf{k}\mathbf{k}'}\) (black) for incoming momentum \(\mathbf{k}\) given by the yellow dot in (a) to out-going momentum given by angle \(\theta_1\) for intra-valley and \(\theta_2\) for inter-valley scattering processes. Phonon mode contributions are shown in colors. Although the electron-phonon coupling strength \(\lambda_\mathbf{k}\) has only slight modulations, the effective potential is strongly dependent on the scattering momentum for a given incoming momentum.}
\label{fig_effectivePotential}
\end{figure*}

Using the parameter values in the preceding subsection, one may calculate the electron-phonon coupling strength \(\lambda\) as function of the chemical potential \(\mu\). This is shown in Fig.~\ref{fig_lambda}(a), with contributions from the four in-plane phonon modes shown in color. The parameter \(\lambda\) incorporates both the strength of the effective potential at the Fermi surface and the density of states. Since the latter has a very systematic variation with the chemical potential, \(\lambda\) and the electronic density of states have similar profiles. In the low doping regime, the optical phonon modes, and the highest energy mode in particular, dominate the electron-phonon coupling strength completely. Fig.~\ref{fig_lambda}(b) shows the angular dependence of \(\lambda_\mathbf{k}\) on the Fermi surface for various dopings. As shown also in the inset, the Fermi surface anisotropy is increasing with doping, reaching values of order 2~\% close to the van Hove singularity.  

The effective potential \(V_{\mathbf{k}\mathbf{k}'}^\mathrm{ph-m.}\) is shown in Fig.~\ref{fig_effectivePotential}(a) for incoming momentum \(\mathbf{k}\) at various outgoing momenta \(\mathbf{k}'\). The potential is attractive in a finite region around the Fermi surface corresponding to the energy of the incoming momentum, and becomes repulsive when the kinetic energy transfer exceeds the phonon energy scale. 

The potential has contributions from the four in-plane phonon modes, and these contributions are shown in Fig.~\ref{fig_effectivePotential}(b) for incoming momentum as indicated in Fig. \ref{fig_effectivePotential}(a). The size of the region with attractive interaction is determined by the energy of the relevant phonon mode. The optical high-energy phonon modes therefore give the largest attractive Brillouin zone area. The effective potential for intra- and inter-valley scattering processes on the Fermi surface is shown in Fig.~\ref{fig_effectivePotential}(c). Comparing the effective potential contribution from the various in-plane phonon modes on the Fermi surface reveals that the high-energy phonon modes corresponding to high mode index or large quasimomentum scattering also give rise to a stronger attractive potential at the important Fermi surface than their low-energetic counterparts.

\subsection{Solutions of the gap equation}

\begin{figure*}[htb]
   \centering
    \includegraphics[width=0.9\textwidth]{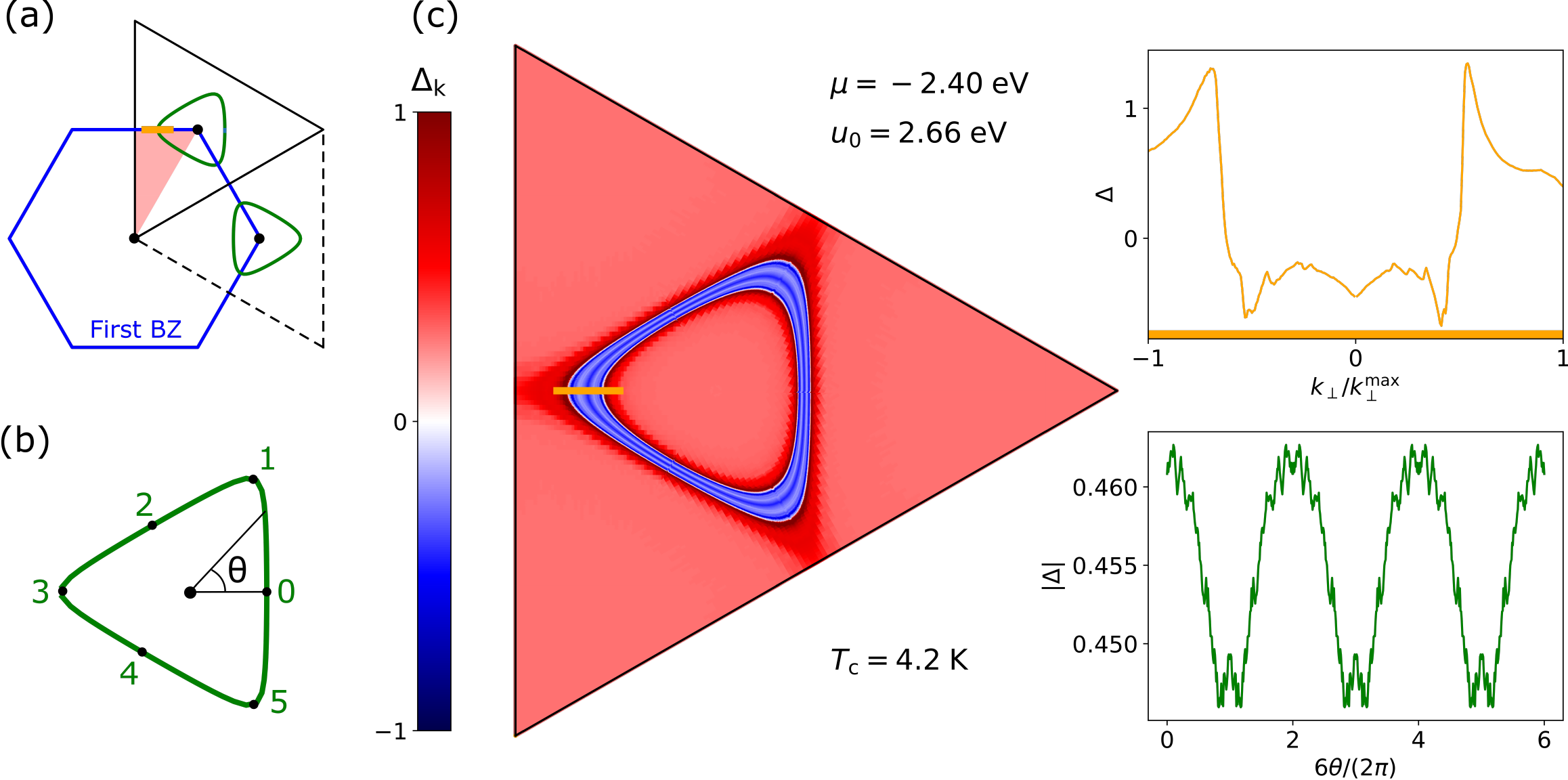}
    \caption{Typical superconducting gap structure at the indicated chemical potential \(\mu\) just below the superconducting transition at critical temperature \(T_c\). (a) Hexagonal Brillouin zone of the triangular Bravais lattice in blue. The rhombus (black) contains an equivalent set of quasimomenta. The green contours indicate the Fermi surface, and the short orange line is perpendicular to the Fermi surface. (b) Position on the Fermi surface is specified with the angle \(\theta\). (c) Gap structure around the point \(K'\) in color for the given doping and on-site repulsion. The insets show the gap structure perpendicular to (orange) and along (green) the Fermi surface. }
    \label{fig_numResults}
\end{figure*}

\begin{figure*}[htb]
    \centering
    \includegraphics[width=0.85\textwidth]{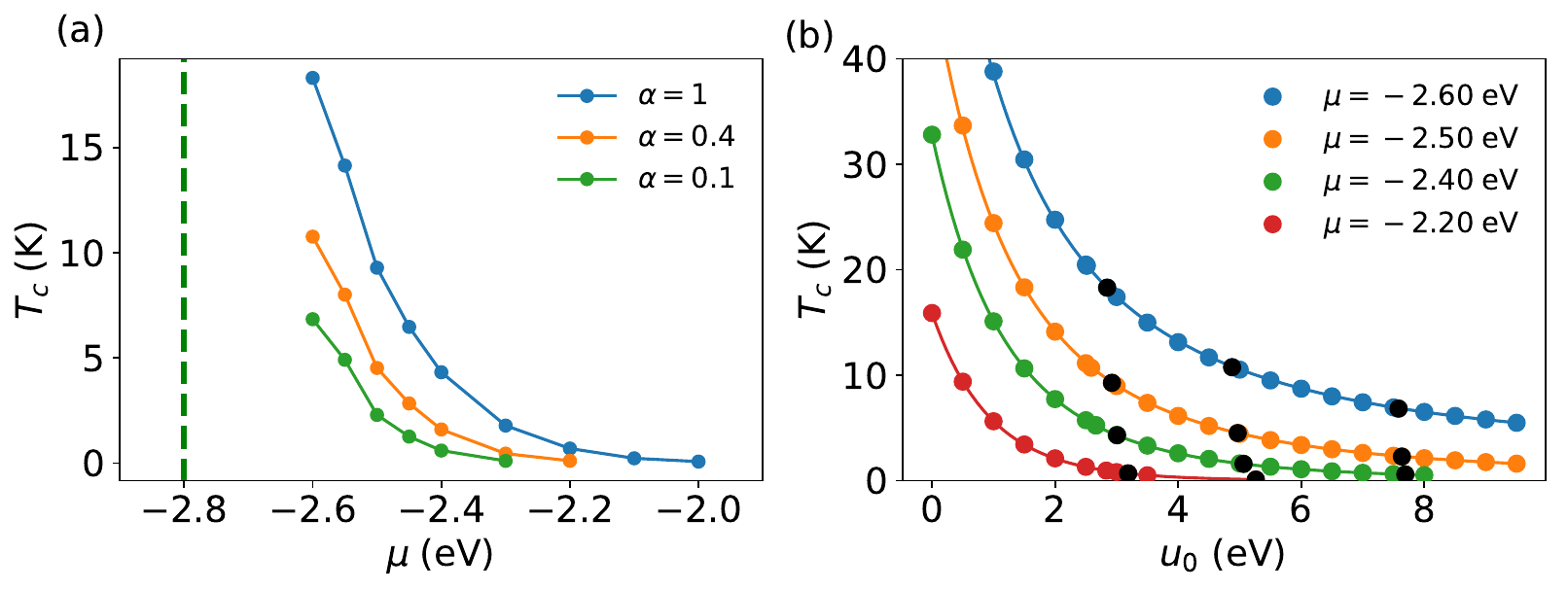}
    \caption{Critical temperature for the superconducting transition. (a)  Critical temperature as function of doping with Hubbard repulsion given by Eq. \eqref{eq_onsiteHubbard} for various polarization strengths parametrized by \(\alpha\). The green dashed line indicates the van Hove singularity. (b) Critical temperature as function of on-site interaction \(u_0\). The calculated data points are fitted to the simple functional form (line) that follows from the Morel-Anderson model. The black points correspond to points in (a).}
    \label{fig_tempDependence}
\end{figure*}

To contain the divergences of the effective electronic potential, we introduce an energy cutoff \(\Lambda = 6 \; \mathrm{eV}\) in the potential. 
Solving the linearized self-consistent equation \eqref{eq_gapEquation} in the full Brillouin zone as discussed in Appendix \ref{sec_appendix_numericalDetails}, we obtain the gap structure at the critical temperature \(T_c\) for which the superconducting instability occurs. This is shown in Fig.~\ref{fig_numResults}, where the superconducting gap at a given point is given by color. 
The gap equation solution shows that the gap has a given sign within the attractive region of the Brillouin zone for incoming momenta at or close to the Fermi surface. Outside this region, the gap changes sign, and subsequently decays to a roughly constant value far away from the Fermi surface. Furthermore, the gap has modulations of the same order as \(\lambda_\mathbf{k}\) along the Fermi surface.

The critical temperature is shown as function of doping in Fig.~\ref{fig_tempDependence}(a). As expected, the critical temperature increases rapidly with increasing doping due to the increasing electron-phonon coupling strength.

The presence of Coulomb interaction decreases the critical temperature significantly. This is shown in Fig.~\ref{fig_tempDependence}(b), which shows the dependence of the critical temperature on the on-site Coulomb repulsion strength \(u_0\). The data points from the solution of the gap equation have been fitted to the simple functional form that is expected from the Morel-Anderson model~\cite{Morel1962}, as discussed in Appendix \ref{sec_appendix_Morel-Anderson}.

\section{Discussion of graphene results}\label{sec_discussion}

In conventional superconductors, the effect of a Coulomb interaction is small, and the quantitative effect on the critical temperature can be incorporated through renormalization~\cite{Morel1962, Alexandrov2003} of the electron-phonon coupling strength \(\lambda\) in the simple BCS result  \(k_B T_c \approx \hbar \omega_D \exp(-1/\lambda)\) according to \(\lambda \rightarrow \lambda - \mu^*\), where

\begin{equation}
\mu^* = \frac{N_0 u}{1 + N_0 u \ln (W/\hbar \omega_D)}.
\end{equation}

\noindent Here, \(u\) is the constant repulsive interaction strength that is added on top of the attractive interaction close to the Fermi surface, \(N_0\) is the density of states at the Fermi surface, \(W\) is the electron band width, and \(\omega_D\) the Debye frequency. For strong Coulomb repulsion, the renormalization is suppressed down to values of \(1/\ln (W/\hbar \omega_D)\), so that Cooper pair formation is possible despite the Coulomb repulsion being much stronger than the attraction at the Fermi surface. 

In the graphene case, simple estimates for the renormalization \(\mu^*\) give a value of \(0.2\) in the presence of strong Coulomb interaction. This is larger than, but not very far away from, estimates~\cite{Si2013, Margine2014} based on the long wavelength limit, arriving at 0.10-0.15. Since the simple Morel-Anderson model predicts the absence of superconductivity for \(\mu^* \geq \lambda\) and we expect to be quite close to this situation, we would expect the Coulomb interactions to have a dramatic effect on the critical temperature of the superconducting transition. Our detailed solution of the gap equation in the presence of Coulomb interaction confirms this picture. Although boosting the electron-phonon coupling \(\lambda\) would be essential for realizing superconductivity in graphene or graphene-like materials, within the realistic regime for \(\lambda\), the repulsive Coulomb interaction also has to be taken into account explicitly. 

Calculations of the critical temperature are notoriously unreliable. On the other hand, the Fermi surface structure of the gap calculated in this paper should give reasonable estimates for the ${\bf k}$-space modulation of the gap on the Fermi-surface. The modulations we find within our methodology are similar and of the same order as in Ref.~\onlinecite{Margine2014}.
The modulations are small, but could in principle be measured by ARPES.

In our calculations, we have considered the electron and phonon band structures of pristine graphene. The presence of intercalant atoms may affect the electron band structure and phonon modes significantly, and this would be dependent on the method chosen to dope graphene~\cite{Hwang2014, Margine2014}. To understand why realizing superconductivity in graphene is so challenging, it is nevertheless useful to study superconductivity based on the intrinsic phonon modes and electronic properties.

In practice, graphene is often mounted on a substrate. A small substrate coupling can be included in our phonon spectrum analysis by adding an on-site potential quadratic in the displacement. This modifies the phonon spectrum by lifting the low energy modes to finite values. Our analysis clearly indicates that it is primarily the high-energy phonons that are responsible for the superconducting instability. Thus, we do not expect a slight alteration of the low-energy phonon-modes to significantly impact our results. Since the introduction of a substrate may break the \(z \rightarrow -z\) mirror symmetry of the system, the out-of-plane modes could in principle also give some contribution to the effective potential, but we expect this to be a higher order effect in the lattice site deviations.

\begin{figure*}[ht]
    \centering
    \includegraphics[width=0.9\textwidth]{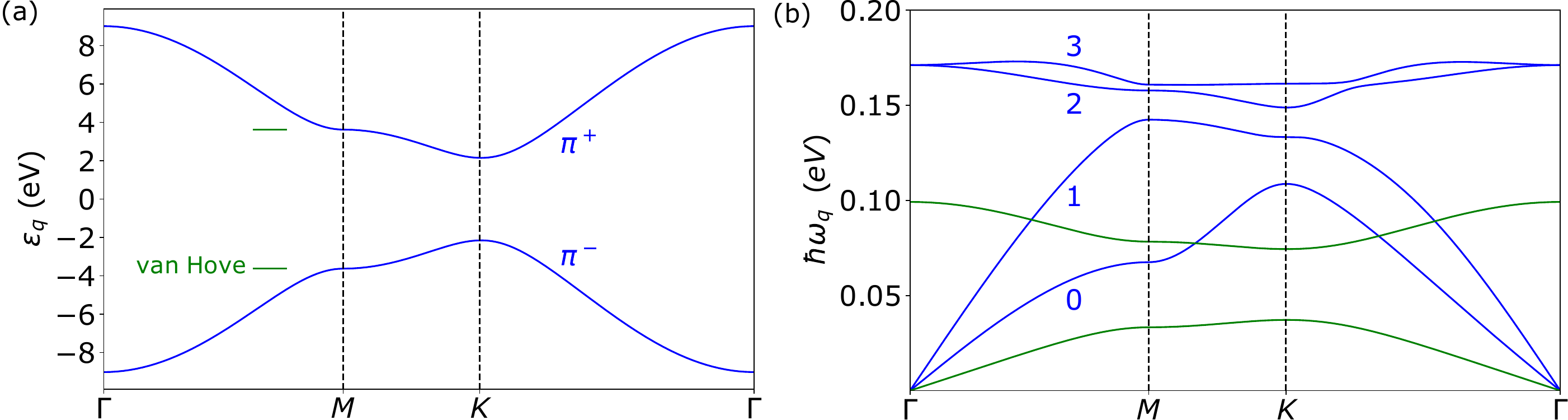}
    \caption{(a) Electron spectrum for the \(\pi\)-band of hexagonal boron nitride (h-BN) in a tight binding hopping model. Contrary to graphene, the band structure is gapped due to sublattice asymmetry.  (b) Phonon spectrum for free-standing h-BN in force constant model. In-plane modes are shown in blue, with out-of-plane modes in green. }

    \label{fig_bnSpectra}
\end{figure*}

\begin{figure}[ht]
\centering
\includegraphics[width=1\columnwidth]{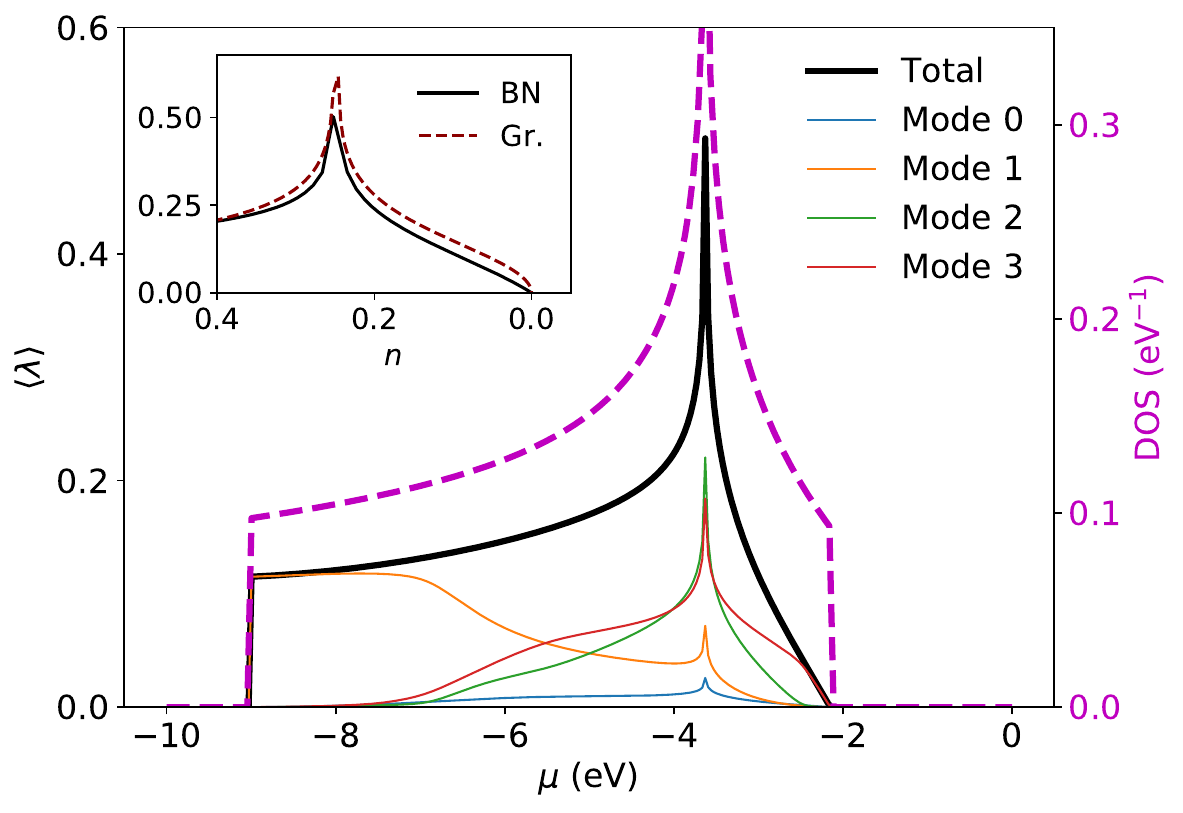}
\caption{Electron-phonon coupling strength \(\lambda\) for boron nitride averaged over the Fermi surface at chemical potential \(\mu\). The contributions to the total electron-phonon coupling (black) from the various phonon modes are shown in colors. The density of states is shown in magenta. A given energy doping \(\mu\) corresponds to a charge doping \(n\) per site, and the inset shows the electron-phonon coupling as a function of this charge doping for graphene and boron nitride.  }
\label{fig_electronPhononCouplingBN}
\end{figure}

\section{Boron nitride}\label{sec_boronNitride}

So far, we have only considered graphene, but our methodology can easily be carried over to other graphene-like materials.
In particular, we consider hexagonal boron nitride (h-BN), which is a two-dimensional material very similar to graphene, but where the atoms on the two different sublattices are boron and nitrogen. The associated sublattice symmetry breaking opens a gap in the electronic spectrum, and in this section, we discuss how this affects the electron-phonon coupling. 

Due to the sublattice symmetry breaking of boron nitride, the electron tight binding model in Eq. \eqref{eq_tightBinding} has to be modified by the addition of a sublattice asymmetric potential term 

\begin{equation}
H_\mathrm{imb} = \frac{\Delta_\mathrm{BN}}{2 } \left( \sum_{i \in A} c_i^\dagger c_i - \sum_{j \in B} c_j^\dagger c_j \right).
\end{equation}

\noindent The resulting electron band structure is shown in Fig.~\ref{fig_bnSpectra}(a), where \(t_1=2.92 \; \mathrm{eV}\) and \(\Delta_\mathrm{BN}=4.3 \; \mathrm{eV}\)~\cite{Robertson1984, Shyu2014}. 

For the phonon excitation spectrum, we again use a force constant model as outlined in Appendix \ref{sec_appendix_phonon}. Values for the boron nitride force constants are obtained by fitting the excitation energies at the high symmetry points to values from density functional theory values in Ref.~\onlinecite{Wirtz2003}, as discussed in Appendix \ref{sec_appendix_forceConstants}. The resulting excitation spectrum is shown in Fig.~\ref{fig_bnSpectra}(b).

As in the graphene calculation, the electron-phonon coupling is obtained by Taylor-expanding the hopping element integral in Eq. \eqref{eq_tightBinding}, and the resulting electron-phonon coupling matrix element is similar~\footnote{Due to the rescaling of the deviation discussed in Appendix \ref{sec_appendix_phonon}, the expression for the electron-phonon coupling matrix element \(g_{\mathbf{k}\mathbf{k}'}^{\eta\eta',\nu}\) is modified according to \(e_\nu^D(\mathbf{q}) \rightarrow e_\nu^D(\mathbf{q})/\sqrt{\mu_D}\). 
} to Eq. \eqref{eq_elPhCouplingMatrixElement}. To compare
the boron nitride results with graphene, we set the value of the dimensionless quantity \(\gamma\) to the same value that was used for graphene. All quantities involved in the calculation of the electron-phonon coupling energy scale \(g_0\) are listed in Appendix  \ref{sec_appendixParameterValues}.

Averaging the dimensionless electron-phonon coupling strength \(\lambda_\mathbf{k}\) over the Fermi surface at chemical potential \(\mu\) gives the result shown in Fig.~\ref{fig_electronPhononCouplingBN}. The inset shows the same electron-phonon coupling \(\lambda\) as function of the charge doping \(n\) corresponding to each chemical potential \(\mu\) for both boron nitride and graphene. 

Unlike the graphene electron-phonon coupling strength shown in Fig.~\ref{fig_lambda}, the electron-phonon coupling strength of boron nitride is qualitatively different from the electronic density of states. At the valence band edge, the electron density of states has a discontinuous jump,
whereas \(\lambda\) increases linearly. Due to the direct onset of a large density of states in boron nitride, it is tempting to assert that even small charge dopings could quickly give rise to appreciable electron-phonon coupling strengths. This is not the case. The electron-phonon coupling matrix element \(|g_{\mathbf{k}\mathbf{k}'}^{--,\nu}|^2\) in Eq. \eqref{eq_elPhCouplingMatrixElement} also plays an essential role for the overall value of the electron-phonon coupling strength, and is suppressed when the Fermi surface is small.
As a result of this, graphene and boron nitride have similar electron-phonon coupling strengths at a given charge doping.

In light of these results, we would expect the difficulty of realizing intrinsic phonon-mediated superconductivity in boron nitride to be similar to that for graphene. Furthermore, the importance of the electron-phonon coupling matrix element underlines the importance of treating the electron-phonon coupling in a detailed manner.

\section{Summary}\label{sec_summary}
In summary, we have studied electron-phonon coupling in graphene and hexagonal boron nitride based on an electron tight-binding and a phonon force constant model giving realistic electron and phonon spectra. The ability to tune the relevant system parameters in this detailed model provides a platform for investigating the superconducting properties of graphene and graphene-like systems. 

In graphene, our results indicate that superconductivity may be possible at sufficiently large doping. We have identified the phonon modes which couple most strongly to $\pi$-band electrons, which are the electronic states of most relevance for realistic doping levels in graphene. These modes are the high-energy in-plane phonon modes.  Solving the gap equation assuming singlet pairing, we find the critical temperature and the superconducting gap structure in the Brillouin zone. The gap has small modulations along the Fermi surface, but is surprisingly uniform even for highly anisotropic Fermi surfaces. Introducing the Coulomb interaction gives a dramatic suppression in the critical temperature, in contrast with the moderate reduction in most normal superconductors. We understand this in terms of the Morel-Anderson model, where the calculated electron-phonon coupling strength and estimates for the renormalization are of the same order. Enhancing the electron-phonon coupling strength is important to realize phonon-mediated superconductivity in monolayer graphene, but the effect of the Coulomb interaction also has to be discussed in detail. 

Motivated by the direct onset of a large density of states in the gapped hexagonal boron nitride, we also calculate the dimensionless electron-phonon coupling there within the same framework. In spite of the large density of states, however, the charge doping required to obtain a sizeable electron-phonon coupling is similar to the doping required in graphene since the electron-phonon coupling matrix element is suppressed due to the small Fermi surface at small charge doping.  

\section{Acknowledgements}

\noindent We thank H. G. Hugdal, E. Erlandsen, V. Saroka, B. Hellsing, and T. Frederiksen for valuable discussions. The Delaunay triangulation procedure was performed using the C++ library delaunator-cpp \cite{Bilonenko2018} by V. Bilonenko, copyrighted with an MIT license (2018).  We acknowledge financial support from the Research Council of Norway Grant No. 262633 ``Center of Excellence on Quantum Spintronics''. A.S. and J.W.W. also acknowledge financial support from the Research Council of Norway Grant No. 250985, ``Fundamentals of Low-dissipative Topological Matter''.

\appendix

\section{Electron band structure}\label{sec_appendix_electron}

To calculate the graphene electron band structure, we use the nearest neighbour tight binding Hamiltonian~\cite{CastroNeto2009},

\begin{equation}
H_\mathrm{el} = 
- t_1 \sum_{\langle ij \rangle, \sigma} \left( c_{i\sigma}^\dagger c_{j\sigma} + \mathrm{h.c.} \right),
\end{equation}

\noindent as our starting point. By introducing the Fourier transformed operators, this model becomes

\begin{equation}
H_\mathrm{el} = \sum_{\mathbf{k},\sigma} 
\begin{pmatrix} c_{\mathbf{k}\sigma A}^\dagger & c_{\mathbf{k}\sigma B}^\dagger \end{pmatrix}
M_\mathrm{k} 
\begin{pmatrix} c_{\mathbf{k}\sigma A} \\ c_{\mathbf{k}\sigma B} \end{pmatrix},
\end{equation}

\noindent where the matrix \(M_\mathbf{k}\) is given by

\begin{equation}
M_\mathbf{k} = 
\begin{pmatrix}
0 & -t_1 \sum_{\pmb{\delta}_A} e^{i\mathbf{k} \cdot \pmb{\delta}_A} \\
-t_1 \sum_{\pmb{\delta}_A} e^{-i\mathbf{k} \cdot \pmb{\delta}_A} & 0 
\end{pmatrix},
\end{equation}

\noindent and \(\pmb{\delta}_A\) are the nearest neighbour vectors from sublattice \(A\) to sublattice \(B\). Diagonalizing this matrix, we get eigenvectors \(F_{D\eta}(\mathbf{k})\) for the two eigenvalues \(\epsilon_{\mathbf{k}\eta}\) corresponding to the two \(\pi\)-bands, where \(\eta\) is the band index. Thus, the \(D\)-sublattice Fourier mode is given by

\begin{equation}
c_{\mathbf{k}\sigma D} = \sum_{\eta} F_{D\eta} (\mathbf{k}) c_{\mathbf{k}\sigma\eta},
\end{equation}

\noindent where \(\eta\) denotes the band and an eigenvector of the matrix \(M_\mathbf{k}\). This provides the definition of the factors \(F_{D\eta}(\mathbf{k})\) appearing in the main text.

\section{Phonon model diagonalization}\label{sec_appendix_phonon}

The phonon dispersion relation calculation in this paper follows Refs.~\onlinecite{Falkovsky2007, Falkovsky2008}, where the phonon excitation spectrum is calculated for graphene. We take the same approach, and use a force constant model with up to third nearest neighbour interactions to calculate the dispersion relations for graphene and boron nitride. Since boron nitride has a broken sublattice symmetry, we have to account for the different sublattice masses, and the intersublattice force constants become sublattice dependent. In this appendix, we discuss how the force constant model can be diagonalized, leaving the discussion of the force constants and their symmetry relations to Appendix \ref{sec_appendix_forceConstants}.

We write the phonon Hamiltonian in the form

\begin{equation}
H_\mathrm{ph} = \sum_j \frac{\mathbf{P}_j^2}{2M_j}  + \frac{1}{2} \sum_{i,j} \sum_{\mu\nu} \Phi_{\mu\nu}^{\kappa_i \kappa_j} (\pmb{\delta}_{ij}) u_{i\mu}^{\kappa_i}  u_{j\nu}^{\kappa_j} ,
\end{equation}

\noindent where \(i, j\) are lattice site indices on the honeycomb lattice, \(\kappa_i, \kappa_j\) are the corresponding sublattices, \(\mu, \nu\) are Cartesian indices, and \(u_{i\mu}^{\kappa_i}\) is the deviation of site \(i\) on the sublattice \(\kappa_i\) (uniquely determined by \(i\)) in direction \(\mu\). The deviation coupling constants are \(\Phi_{\mu\nu}^{\kappa_i \kappa_j}(\pmb{\delta}_{ij})\). In the kinetic term, \(\mathbf{P}_j\) is the momentum of the particle at site \(j\), and \(M_j\) is the mass. 

We next express the phonon Hamiltonian in terms of uncoupled harmonic oscillators. To do this, we first symmetrize the sublattice sectors of the kinetic term. Introducing effective mass \(\tilde{M}=\sqrt{M_A M_B}\) and relative masses \(\mu_D\) given by \(M_D = \mu_D \tilde{M}\), we introduce rescaled deviations and momenta

\begin{equation}
\tilde{\mathbf{P}}^D = \mathbf{P}_D / \sqrt{\mu_D} 
\qquad
\tilde{\mathbf{u}}^D = \mathbf{u}^D \sqrt{\mu_D},
\end{equation}

\noindent where the rescaling of the deviations is chosen to retain the canonical commutation relations \([u_{i\mu}, P_{j\nu}] = i\hbar \delta_{ij} \delta_{\mu\nu}\). 
To proceed, we rewrite the Hamiltonian in Fourier space, obtaining 

\begin{equation}
H_\mathrm{ph}
= \frac{1}{2 \tilde{M}} \sum_{\kappa,\mathbf{q}} {\tilde{\mathbf{P}}_\mathbf{-q}^\kappa \tilde{\mathbf{P}}_\mathbf{q}^\kappa} + \frac{1}{2} \sum_{\kappa \kappa'} \sum_{\mu\nu} \sum_\mathbf{q} D^{\kappa\kappa'}_{\mu\nu} (\mathbf{q} )
\tilde{u}_{-\mathbf{q},\mu}^\kappa  \tilde{u}_{\mathbf{q}\nu}^{\kappa'},
\end{equation}

\noindent where \(\kappa, \kappa'\) are sublattice indices and the matrix elements \(D_{\mu\nu}^{\kappa\kappa'}(\mathbf{q})\) are given by

\begin{equation}
D^{\kappa\kappa'}_{\mu\nu} (\mathbf{q} ) = \frac{1}{\sqrt{\mu_\kappa \mu_{\kappa'}} }\sum_{j \in \kappa'} \Phi_{\mu\nu}^{\kappa\kappa'} (\pmb{\delta}_j^\kappa) e^{i\mathbf{q} \cdot \pmb{\delta}_j^{\kappa} },
\label{eq_phononMatrixFourier}
\end{equation}

\noindent where \(\pmb{\delta}_j^\kappa\) is the vector from a lattice site on sublattice \(\kappa\) to lattice site \(j\) on sublattice \(\kappa'\).

Using the symmetries of the system, as discussed further in Appendix \ref{sec_appendix_forceConstants}, the number of independent real space coupling constants can be reduced drastically. Here, we only point out the effect of the mirror symmetry under \(z\rightarrow -z\). Considering the lattice deviation coupling term in the phonon Hamiltonian, this symmetry implies that there cannot be any coupling between the in-plane and the out-of-plane modes, and hence that the phonon eigenmodes are either purely in-plane or out-of-plane.  The potential energy term can thus be written in the form \(V_\mathrm{ph} = V_\mathrm{ph}^{z} + V_\mathrm{ph}^{xy}\), where

\begin{equation}
\begin{aligned}
V_\mathrm{ph}^{xy} &= \frac{1}{2} \sum_{\mathbf{q}} 
(\tilde{\underline{u}}^{xy}_{\mathbf{q}})^\dagger M_\mathbf{q}^{xy} \tilde{\underline{u}}^{xy}_{\mathbf{q}} \\
V_\mathrm{ph}^{z} &= \frac{1}{2} \sum_{\mathbf{q}} 
(\tilde{\underline{u}}_{\mathbf{q}}^z)^\dagger M_\mathbf{q}^{z} \tilde{\underline{u}}_{\mathbf{q}}^z
\end{aligned}
\end{equation}

\noindent and the deviations \(\tilde{\underline{u}}_\mathbf{q}\) are given by

\begin{equation}
\begin{aligned}
\tilde{\underline{u}}_\mathbf{q}^{z}
&=
\begin{pmatrix} 
\tilde{u}^{A}_{\mathbf{q},z} & 
\tilde{u}^{B}_{\mathbf{q},z} 
\end{pmatrix}^T \\
\tilde{\underline{u}}_\mathbf{q}^{xy}
&=
\begin{pmatrix} 
\tilde{u}^{A}_{\mathbf{q},x} & 
\tilde{u}^{A}_{\mathbf{q},y} & 
\tilde{u}^{B}_{\mathbf{q},x} & 
\tilde{u}^{B}_{\mathbf{q},y}
\end{pmatrix}^T.
\end{aligned}
\end{equation}

The matrices \(M_\mathbf{q}^{z}\) and \(M_\mathbf{q}^{xy}\) are \(2\times 2\) and \(4 \times 4\) matrices, and the matrix elements for graphene are given in Ref.~\onlinecite{Falkovsky2007}. For the boron nitride case, similar expressions are derived by inserting values for the coupling constants using the symmetry relations and force constants in Appendix \ref{sec_appendix_forceConstants}.

To obtain a system of uncoupled harmonic oscillators, we introduce a new basis \(v_\mathbf{q}^\nu\) given by

\begin{equation}
\tilde{u}_{\mathbf{q}\mu}^D = \sum_\nu [\mathbf{e}^D_\nu(\mathbf{q})]_\mu v_\mathbf{q}^\nu
\end{equation}

\noindent in which the phonon potential energy is diagonal. Here, \([e_\nu(\mathbf{q})]_\mu\) is given by the eigenvectors of \(M_\mathbf{k}\), \(\nu\) is an eigenvector label, \(\mathbf{e}_\nu^D (\mathbf{q})\) is the phonon polarization vector on sublattice \(D\) at quasimomentum \(\mathbf{q}\), and the index \(\mu\) picks out a Cartesian component. This relation provides a definition for the phonon polarization vectors occurring in the electron-phonon coupling in the main text.

Since the kinetic term remains diagonal in the new basis, the system is reduced to a system of uncoupled harmonic oscillators, from which we obtain~\cite{Kittel1987} the excitation spectrum through \(\omega_{\mathbf{q}\nu}^2 = d_{\mathbf{q}\nu}/\tilde{M}\), where \(d_{\mathbf{q}\nu}\) are the eigenvalues of \(D(\mathbf{q})\). 

In our paper, the phonon spectrum and associated polarization vectors \(\mathbf{e}_\nu^D(\mathbf{q})\) are determined using numerical diagonalization. At the high symmetry point \(K\), one may derive reasonably simple expressions for the eigenfrequencies.

\section{Force constants and symmetries}\label{sec_appendix_forceConstants}

The discussion in this Appendix is a generalization of the graphene force constant model in Refs.~\onlinecite{Falkovsky2007, Falkovsky2008} to the case of honeycomb lattices without sublattice symmetry. We provide an overview of how the symmetries of the system impose relations between the force constants in the model, and determine the  force constants by fitting the force constant dispersion relation to density functional theory results in Ref.~\onlinecite{Wirtz2003}.

\subsection{Chiral basis and double counting}

The phonon Hamiltonian can be written in the form

\begin{equation}
H_\mathrm{ph} = \sum_j \frac{\mathbf{P}_j^2}{2M_j}  + \frac{1}{2} \sum_{i,j} \sum_{\mu\nu} \Phi_{\mu\nu}^{\kappa_i \kappa_j} (\pmb{\delta}_{ij}) u_{i\mu}^{\kappa_i}  u_{j\nu}^{\kappa_j} \nonumber,
\end{equation}

\noindent where \(\sum_{i} \) denotes the sum over all lattice sites on the honeycomb lattice,  and all bonds \((i,j)\) are being double counted. To symmetrize these contributions, we may therefore impose

\begin{equation}
\Phi_{\mu\nu}^{\kappa_i \kappa_j}(\pmb{\delta}_{ij})
= 
\Phi_{\nu\mu}^{\kappa_j \kappa_i}(\pmb{\delta}_{ji}),
\label{eq_symRel_doubleCounting}
\end{equation}

\noindent where the indices \(\mu,\nu\) are initially considered to be Cartesian. We may however also introduce the chiral basis

\begin{align}
\xi = x + i y  \qquad \eta = x - i y,
\end{align}

\noindent so that \(\mu, \nu \in \{\xi, \eta, z\}\). Under rotation with angle \(\phi\) around the z-axis, the new coordinates do not mix, and transform according to

\begin{equation}
\xi \rightarrow \xi e^{i\phi}
\qquad
\eta \rightarrow \eta e^{-i\phi}.
\end{equation}

In terms of the old coupling coefficients, the coefficients for the deviations in the new basis are given by

\begin{equation}
\begin{aligned}
\Phi_{\xi\xi} &= \left( \Phi_{xx} - \Phi_{yy} - i \Phi_{xy} - i \Phi_{yx} \right)/4 \\
\Phi_{\eta\eta} &= \left( \Phi_{xx} - \Phi_{yy} + i \Phi_{xy} + i \Phi_{yx} \right)/4 \\
\Phi_{\xi\eta} &= \left( \Phi_{xx} + \Phi_{yy} + i \Phi_{xy} - i \Phi_{yx} \right)/4 \\
\Phi_{\eta\xi} &= \left( \Phi_{xx} + \Phi_{yy} - i \Phi_{xy} + i \Phi_{yx} \right)/4.
\end{aligned}
\label{eq_symRel_cartesianCoefficients}
\end{equation}

\noindent Now, both deviations and coupling constant are in general complex. 

\subsection{Force constant symmetry relations}

The hexagonal boron nitride system has infinitesimal translation symmetry, Bravais lattice translation symmetry, infinitesimal rotation symmetry, lattice \(C_3\) rotation symmetry, \(\sigma_z\) mirror symmetry, and, with the choice of lattice orientation indicated in Fig. \ref{fig_nnVectors}, \(\sigma_x\) mirror symmetry. We use these symmetries to reduce the number of independent coupling coefficients. 

\subsubsection{Translation symmetries}

From translation symmetry with a Bravais lattice vector \(\mathbf{a}\), it follows trivially, as already indicated by the force constant notation, that 

\begin{equation}
\Phi_{\mu\nu}^{\kappa_i \kappa_j} (\pmb{\delta}_{ij})
= 
\Phi_{\mu\nu}^{\kappa_i \kappa_j} (\pmb{\delta}_{i+\mathbf{a}, j+\mathbf{a}}).
\end{equation}

Due to the infinitesimal translation symmetry of a single graphene sheet under \(u_{i\mu}^{\kappa_i} \rightarrow u_{i\mu}^{\kappa_i} + a_\mu\), it furthermore follows that 

\begin{equation}
\sum_j \Phi_{\mu\nu}^{\kappa_i \kappa_j} (\pmb{\delta}_{ij}) = 0.
\end{equation}

\noindent Following Refs.~\onlinecite{Falkovsky2007, Falkovsky2008}, we call this the stability condition, and use it to determine the local force constants with \(\pmb{\delta}_{ij} = 0\). 

Although infinitesimal lattice translation symmetry holds for a freestanding graphene sheet, it breaks down if the monolayer sheet is placed on a substrate. This would give rise to additional free parameters through the force constants \(\Phi_{\mu\nu}^{DD}(0)\).  

\subsubsection{Rotation symmetries}

Application of the \(C_3\)-symmetry under 3-fold rotations \(R_3\) gives force constant relations

\begin{equation}
\begin{aligned}
\Phi_{\xi\xi}^{\kappa_i \kappa_j} (R_3 \pmb{\delta}_{ij}) 
&= \Phi_{\xi\xi}^{\kappa_i \kappa_j} (\pmb{\delta}_{ij}) e^{+i 2\pi/3} \\
\Phi_{\eta\eta}^{\kappa_i \kappa_j} (R_3 \pmb{\delta}_{ij}) 
&= \Phi_{\eta\eta}^{\kappa_i \kappa_j} (\pmb{\delta}_{ij}) e^{-i 2\pi/3},
\end{aligned}
\end{equation}

\noindent whereas \(\Phi_{\mu\nu}^{\kappa_i \kappa_j} (R_3 \pmb{\delta}_{ij}) = \Phi_{\mu\nu}^{\kappa_i \kappa_j} (\pmb{\delta}_{ij}) \) if \(\mu\) and \(\nu\) are not equal chiral in-plane components, as in the two cases listed above.

We also note that the infinitesimal rotation symmetry does not give restrictions on the force constants in addition to the ones we have already discussed.

\subsubsection{Mirror symmetries and complex conjugation}

The mirror symmetry \(\sigma_z\) implies that there cannot be any coupling between the in-plane and the out-of-plane deviations, i.e.

\begin{equation}
\Phi_{\xi z} = \Phi_{\eta z} = \Phi_{z \xi} = \Phi_{z\eta} = 0.
\end{equation}

\noindent As already discussed in Appendix \ref{sec_appendix_phonon}, this completely decouples the in-plane and the out-of-plane phonon modes.

The \(\sigma_x\) mirror symmetry implies

\begin{equation}
\Phi_{\mu\nu}^{\kappa_i \kappa_j} ( \pmb{\delta}_{ij} )
= 
\Phi_{\bar{\mu}\bar{\nu}}^{\kappa_i \kappa_j} ( \sigma_x \pmb{\delta}_{ij} ),
\end{equation}

\noindent where \(\bar{\xi} = \eta\), \(\bar{\eta}=\xi\) and \(\bar{z} = z\).

Finally, we note that the requirement of a real potential gives the relation

\begin{equation}
\Phi_{\mu\nu}^{\kappa_i \kappa_j} (  \pmb{\delta}_{ij} )
=
\Phi_{\bar{\mu}\bar{\nu}}^{\kappa_i \kappa_j} (  \pmb{\delta}_{ij} )^*,
\end{equation}

\noindent and this can be combined with the above mirror symmetry \(\sigma_x\) to obtain  

\begin{equation}
\Phi_{\mu\nu}^{\kappa_i \kappa_j} (  \pmb{\delta}_{ij} )
=
\Phi_{{\mu}{\nu}}^{\kappa_i \kappa_j} ( \sigma_x \pmb{\delta}_{ij} )^*.
\label{eq_symRel_complexSigma_x}
\end{equation}

\noindent For the case of neighbour vectors parallel to the y-axis, invariance of the neighbour vector under the mirror symmetry \(\sigma_x\) implies that the coupling constant has to be real.

\subsection{Boron nitride force constants}

\begin{figure}
    \centering
    \includegraphics[width=0.5\columnwidth]{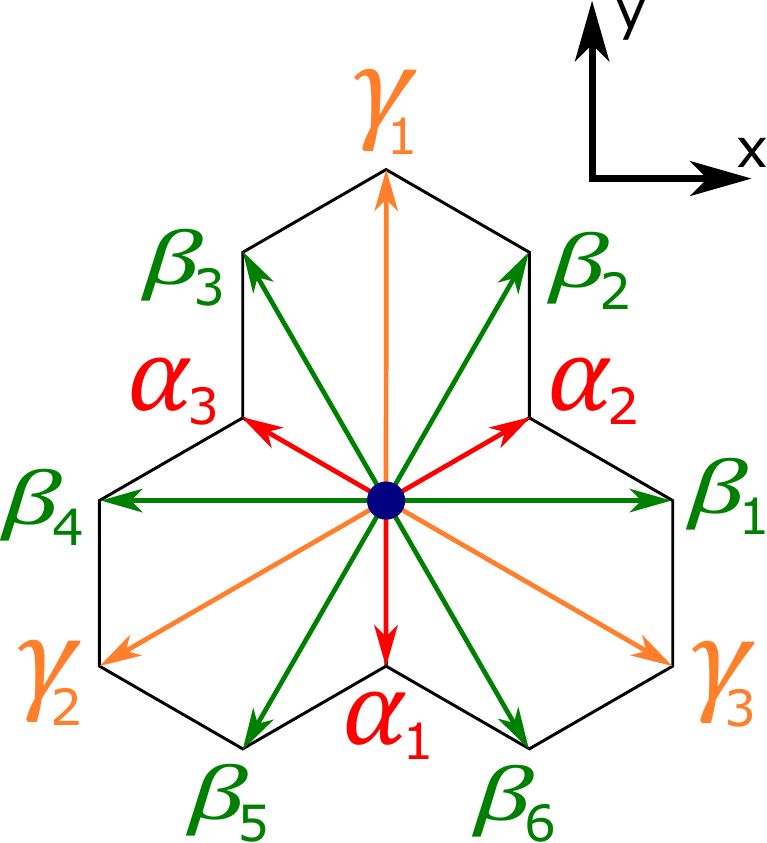}
    \caption{Labelling of vectors to neighbouring sites on the honeycomb lattice.}
    \label{fig_nnVectors}
\end{figure}

\begin{table}
\caption{Force constants for graphene and boron nitride phonons up to next-to-nearest neighbour for graphene and third nearest neighbour for boron nitride. The graphene force constants are taken from Ref.~\onlinecite{Falkovsky2007}. The tabulated values give \(\Phi/\tilde{M}\) in spectroscopic units of \(10^5 \; \mathrm{cm}^{-2}\), related to frequency through factors of \(2\pi c\), where \(c\) is the speed of light.}
\label{tab_forceConstants}
\begin{center}
\begin{tabular}{c c c c c c}
\hline
Parameter & Coupling &  \(\mathbb{R}\)/\(\mathbb{C}\)  &  Graphene & h-BN \\
\hline
\(\alpha\)  & \(\Phi_{\xi\eta}^{AB}(\pmb{\alpha}_1) \) & \(\mathbb{R} \) & \(-4.046\) & \(-3.15\) \\
\(\beta\)  & \(\Phi_{\xi\xi}^{AB}(\pmb{\alpha}_1) \) & \(\mathbb{R} \) &  1.107& \(1.69\) \\
\(\gamma_A\)  & \(\Phi_{\xi\eta}^{AA}(\pmb{\beta}_1) \) & \(\mathbb{C} \) & \(-0.238 \)& \(-0.32 + 0.05i\) \\
\(\gamma_B^*\) & \(\Phi_{\xi\eta}^{BB}(\pmb{\beta}_1) \) & \(\mathbb{C} \)  & \(-0.238\)& \(-0.36 - 0.07i\) \\
\(\delta_A\)  & \(\Phi_{\xi\xi}^{AA}(\pmb{\beta}_1) \) & \(\mathbb{R} \) & \(-1.096\) & \(-0.68\) \\
\(\delta_B\)  & \(\Phi_{\xi\xi}^{BB}(\pmb{\beta}_1) \) & \(\mathbb{R} \) & \(-1.096\) & \(-0.66\)\\
\(\alpha'\) &  \(\Phi_{\xi\eta}^{AB}(\pmb{\gamma}_1) \) & \(\mathbb{R} \) & - & \(0.00\) \\
\(\beta'\) &   \(\Phi_{\xi\xi}^{AB}(\pmb{\gamma}_1) \) & \(\mathbb{R} \) & - & \(-0.23\) \\
\hline
\(\alpha_z\)  & \(\Phi_{zz}^{AB}(\pmb{\alpha}_1) \) & \(\mathbb{R} \) & -1.176& -1.06\\
\(\gamma_z^A\)  & \(\Phi_{zz}^{AA}(\pmb{\beta}_1) \) & \(\mathbb{R} \) & 0.190& 0.00\\
\(\gamma_z^B\)  & \(\Phi_{zz}^{BB}(\pmb{\beta}_1) \)& \(\mathbb{R} \) & 0.190 & 0.24\\
\hline
\end{tabular}
\end{center}

\end{table}

Applying the above symmetry relations, the independent force constants in the system are listed in Table \ref{tab_forceConstants} along the bonds illustrated in Fig.~\ref{fig_nnVectors}. The graphene force constants are taken from Ref.~\onlinecite{Falkovsky2007}, and the boron nitride force constants have been obtained by fitting the phonon frequencies at the high symmetry points to density functional theory results in Ref.~\onlinecite{Wirtz2003}. Other force constants in the system can be determined from the force constant symmetry relations listed above. 

Note that, contrary to what Refs.~\onlinecite{Falkovsky2007, Falkovsky2008} claim, the force constants \(\gamma_D\) are in general complex, whereas the other independent force constants up to third-nearest-neighbours, including \(\delta_D\), are real. This can be seen from the symmetry relation in Eq. \eqref{eq_symRel_complexSigma_x} and the double counting symmetrization relation in Eq. \eqref{eq_symRel_doubleCounting}, as well as the mirror symmetry \(\sigma_x\) in combination with the Cartesian component expressions in Eq. \eqref{eq_symRel_cartesianCoefficients}. 

\section{Coulomb interaction model}\label{sec_appendix_coulomb}

The Coulomb interaction in a lattice model such as ours can be modelled with the Hubbard type interaction

\begin{equation}
V^C = u_0 \sum_i n_{i\uparrow} n_{i\downarrow} 
+ \sum_{ ij} u_{ij} n_i n_j,
\end{equation}

\noindent where \(u_{ij}\) are the non-local interaction parameters. In this Appendix, we discuss how one may model the doping dependence of the non-local interaction strength parameters. The doping dependence of the on-site repulsion is discussed in the main text.

The on-site and two nearest neighbour interaction strength parameters were calculated for pristine graphene in Ref.~\onlinecite{Wehling2011} based on density functional theory and the constrained random phase approximation. Any pristine interaction strength parameter can therefore be modelled through the combination of these values and Coulombic decay~\cite{Ulybyshev2013}.

At finite doping, we expect the onset of \(\pi\)-band screening to reduce the interaction coefficients \(u_{ij}(\mu)\). To obtain an estimate for the non-local interaction parameters, one may write

\begin{equation}\label{eq_screenedInteractionCoefficients}
u_{ij}(\mu) = \frac{V^\textrm{sc}_\mu ( r_{ij} ) } {V_0(r_{ij}) } u_{ij}(0), 
\end{equation}

\noindent where \(V_0(r)\) is the potential screened only by the \(\sigma\)-bands and the substrate, and \(V_{\mu}^\textrm{sc}(r)\) is the potential screened also by \(\pi\)-band Dirac electrons. In the long wavelength limit, the screened interaction is~\cite{Kotov2012, Hwang2007}

\begin{equation}
V_\mu^\textrm{sc}(q) = \frac{1}{2 \epsilon_0}  \left(\frac{e^2}{q +q_0} \right),
\end{equation}

\noindent where the Thomas-Fermi momentum \(q_0\) is given by~\cite{Hwang2007}

\begin{equation}
q_0 = \frac{e^2 |\mu|}{\pi \hbar^2 v^2  \epsilon_r \epsilon_0}. 
\end{equation}

\noindent Here, \(v\) is the Dirac cone velocity, and \(\epsilon_r\) a relative permittivity depending on the substrate~\cite{Kotov2012}. Inserting parameter values, we obtain the screening length \(1 /q_0 d = 0.43 \epsilon_r \; {\mathrm{eV}}/|\mu|\).

In real space, the screened interaction takes the form

\begin{equation}
V_\mu^\mathrm{sc}(r) = \frac{1}{4\pi \epsilon_r \epsilon_0 } \left(
\frac{1}{r} - \frac{\pi}{2} q_0 \left[ H_0(q_0 r) - N_0(q_0 r) \right] \right),
\end{equation}

\noindent where \(H_0(x)\) is the Struve function and \(N_0(x)\) the Bessel function of second kind~\cite{Stern1967}. Through asymptotic expansion of the Struve and Bessel functions~\cite{AbramowitzStegun1965}, one may show that the screened potential has long distance behaviour \(V^\mathrm{sc}_\mu(r) \sim 1/r^3\). The screening length \(1/q_0\) determines the crossover point to this rapidly decaying long distance behaviour from the Coulombic small distance behaviour. 

Since the screening length is a small fraction of the lattice constant for significant doping of order \(\SI{2}{\eV}\), we keep only the on-site Hubbard interaction. For this on-site term, Eq. \eqref{eq_screenedInteractionCoefficients} can no longer be used, and as discussed in Sec. \ref{sec_coulomb} of the main paper, we instead use the direct polarization bubble renormalization.


\section{Solving the gap equation}\label{sec_appendix_numericalDetails}

The gap equation is given by

\begin{align}
\Delta_\mathbf{k} 
= 
-\frac{1}{A_{BZ}} 
\int d^2 k' \; \tilde{V}_{\mathbf{k}\mathbf{k}'}^{\mathrm{symm}} \chi_{\mathbf{k}'} \Delta_{\mathbf{k}'} \label{eq_gapEquationIntegral}
\end{align}

\noindent where \(A_{BZ}\) is the Brillouin zone area and we let \(\tilde{V}_{\mathbf{k}\mathbf{k}'}^{\mathrm{symm}} = N_A {V}_{\mathbf{k}\mathbf{k}'}^{\mathrm{symm}}\).

To find a proper solution to the discretized version of this gap equation, it is important to have sufficiently good resolution in the important regions of the Brillouin zone. The factor \(\chi_{\mathbf{k}}\) is peaked around the Fermi surface with a peak width \(\propto T\) and necessitates a good resolution there. Furthermore, good resolution is also required in the regions around the corners of the triangle-like Fermi surface at significant doping. To make sure of this, we select points on a uniform grid in the Brillouin zone, add additional points close to the Fermi surface, and further additional points close to the Fermi surface corners. 

To solve the gap equation, we rewrite the gap equation on the integral form of Eq. \eqref{eq_gapEquationIntegral} in terms of a weighted sum over the points described in the previous paragraph. To find the appropriate weights \(w_\mathbf{k}\), we split the Brillouin zone into triangles \(\{t\}\) with the points \(\{\mathbf{k}\}\) as vertices using Delaunay triangulation. Denote the area of a triangle \(t\) by \(A_t\). The weight of a single point then becomes one third of the sum of the areas of all the triangles that has the point as a vertex, i.e. 

\begin{equation}
w_\mathbf{k} = \sum_{t}  A_t \delta_{\mathbf{k} \in t}/3,
\end{equation}

\noindent where \(\delta_{\mathbf{k} \in t}\) is 1 if \(\mathbf{k}\) is a vertex in the triangle \(t\) and 0 otherwise. 

The gap equation then becomes

\begin{align}
\Delta_\mathbf{k} &= -\frac{1}{A_{BZ}} \sum_{\mathbf{k}'} \tilde{V}_{\mathbf{k}\mathbf{k}'}^{\mathrm{symm}} w_{\mathbf{k}'} \chi_{\mathbf{k}'} \Delta_{\mathbf{k}'}.
\end{align}

The symmetrized potential \(\tilde{V}_{\mathbf{k}\mathbf{k}'}^{\mathrm{symm}}\) is symmetric under the exchange of incoming and outgoing momenta, but to symmetrize the eigenvalue problem in this exchange, we multiply this equation with \(\sqrt{w_\mathbf{k} \chi_\mathbf{k}}\) on both sides, to obtain a gap equation in the form

\begin{equation}
\tilde{\Delta}_\mathbf{k}
= 
\sum_{\mathbf{k}'} M_{\mathbf{k}\mathbf{k}'}(\beta) 
\tilde{\Delta}_{\mathbf{k}'},
\end{equation}

\noindent where we introduced the weighted gap \(\tilde{\Delta}_\mathbf{k} = \sqrt{w_\mathbf{k} \chi_\mathbf{k} } \Delta_\mathbf{k}\), and the matrix

\begin{equation}
M_{\mathbf{k}\mathbf{k}'} = - \frac{1}{A_{BZ}} 
\left( 
\sqrt{w_\mathbf{k} \chi_\mathbf{k}} \tilde{V}_{\mathbf{k}\mathbf{k}'}^{\mathrm{symm}} \sqrt{w_{\mathbf{k}'} \chi_{\mathbf{k}'}}
\right)
\end{equation}

\noindent is symmetric in the interchange of \(\mathbf{k}\) and \(\mathbf{k}'\). 

To reduce the size of the matrix \(M\) and improve computational efficiency, we split the Brillouin zone into small triangles similar to the shaded red triangle in Fig.~\ref{fig_numResults}(a), and assume that the gap takes the same value at corresponding points in all the triangles. The effective potential corresponding to a scattering process within the shaded red triangle is then the sum of contributions for scatterings to outgoing momenta in all the small triangles which correspond to the outgoing momentum within the shaded triangle. This reduction of the problem excludes gap equation solutions without the full symmetry of graphene, but we have checked that we obtain the same solutions by solving the gap equation in the full Brillouin zone.

We now have a matrix eigenvalue problem linear in the eigenvectors and non-linear in the eigenvalue. We find the gap structure at the superconducting instability by determining the smallest \(\beta\), i.e. the largest temperature, for which the largest eigenvalue of \(M_{\mathbf{k}\mathbf{k}'}\) becomes 1. The corresponding eigenvector must be a solution of our eigenvalue problem. The critical temperature \(T_c = 1/\beta_c\) is located using the bisection algorithm.

\section{Parameter values}\label{sec_appendixParameterValues}

\begin{table}[htb]
\begin{center}
\caption{Values for the quantities involved in the calculation of the electron-phonon coupling amplitude strength \(g_0\), where the \(A\)-sublattice of boron nitride is assumed to host boron and the \(B\)-sublattice nitrogen.}
\label{tab_parameterValues}
\begin{tabular}{llll}
\hline
Quantity & Graphene & h-BN & Description\\
\hline
\(d\) & 1.42 \AA & 1.45 \AA & NN-distance \\
\(t_1\) & 2.8 eV & 2.92 eV & Hopping amplitude\\
\(\Delta\) & 0 & 4.30 eV & Band gap\\
\(\hbar \omega_\Gamma\) & 0.20 eV  & 0.17 eV & Phonon energy scale\\
\(\tilde{M}\) & \( 12.0 \;\text{u} \)  & \(12.3 \; \mathrm{u}\) & Effective mass \\
\({\mu_A}\)  & 1 & 0.88 & Relative mass, \(A\) subl. \\
\({\mu_B}\)  & 1 & 1.14 & Relative mass, \(B\) subl.\\
\(\gamma\) & 2.5 & 2.5 & \(-\mathrm{d} \ln t_1 / \mathrm{d} \ln d\) \\
\(m_e\) &   \(5.49 \cdot 10 ^{-4} \;\text{u}\) &  & Electron mass \\
\( 1\;\mathrm{Ry}\) &   13.6 eV &  & Rydberg energy\\
\(a_0\) & 0.53 \AA &  & Bohr radius \\
\(g_0\) & 0.15 eV & 0.16 eV & El-ph coupling scale \\
\hline
\end{tabular}
\end{center}
\end{table}

The parameter values used in the electron tight binding model and the electron-phonon coupling for graphene and boron nitride are listed in Table \ref{tab_parameterValues}. 
The electron-phonon coupling scale \(g_0\) can be written as

\begin{equation}
 g_0 =   \gamma t_1
\sqrt{ \left(\frac{\hbar^2 }{2m_e a_0^2}\right)  \frac{1 }{\hbar \omega_\Gamma} \left( \frac{m_e}{M} \right) \left( \frac{a_0}{d} \right)^2},
\end{equation}

\noindent where \(m_e\) is the electron mass, and \(a_0\) the Bohr radius. This quantity is calculated based on the listed parameter values, and also given in the table.

\section{Morel-Anderson model}\label{sec_appendix_Morel-Anderson}

\noindent The Morel-Anderson model is a simple model describing the effect of a repulsive potential in the entire Brillouin zone on top of an attractive potential in a small region around the Fermi surface giving rise to superconductive pairing \cite{Morel1962}. This model illustrates why there can be a superconducting instability even though the interaction potential is repulsive even close to the Fermi surface. 

In the Morel-Anderson model, one assumes that the potential \(V_{\mathbf{k}\mathbf{k}'}\) occurring in the gap equation takes the form \(V_{\mathbf{k}\mathbf{k}'} = V_{\mathbf{k}\mathbf{k}'}^\mathrm{rep.} + V_{\mathbf{k}\mathbf{k}'}^\mathrm{attr.}\) with repulsive and attractive potentials

\begin{align}
V_{\mathbf{k}\mathbf{k}'}^\mathrm{rep.} = 
\left\{\begin{array}{ll}
        u & \text{for } -W \leq \xi_{\mathbf{k}}, \xi_{\mathbf{k}'}  \leq W  \\
        0 & \text{otherwise} 
\end{array}\right\}
\end{align}

\noindent and 

\begin{align}
V_{\mathbf{k}\mathbf{k}'}^\mathrm{attr.} = 
\left\{\begin{array}{ll}
        -v & \text{for } -\epsilon_{D} \leq \xi_{\mathbf{k}}, \xi_{\mathbf{k}'}  \leq \epsilon_{D} \\
        0 & \text{otherwise} 
\end{array}\right\},
\end{align}


\noindent where \(u,v \geq 0\), \(W\) is the bandwidth cutoff, and \(\epsilon_D = \hbar \omega_D\) represents the size of the region with attractive interactions around the Fermi surface. In the case of phonon-mediated superconductivity, this is the phonon Debye frequency. 

The gap equation for singlet BCS pairing can now be solved by turning the momentum integral into an energy integral, approximating the density of states by the density of states \(N_F\) at the Fermi surface, and assuming the gap to take on two different constant values close to (\(|\xi_\mathbf{k}| \leq \epsilon_D\)) and far away from (\(|\xi_\mathbf{k}| > \epsilon_D\)) the Fermi surface.

This gives a critical temperature given by

\begin{equation}
k_B T_c = 1.14 \; \epsilon_D \exp \left\{ -\frac{1}{\lambda - \mu^*} \right\},
\end{equation}

\noindent where \(\lambda= N_F v\) is the potential strength of the attractive potential, and

\begin{equation}
\mu^* = \frac{N_F {u}}{1 +  N_F {u} \ln (W/\hbar \omega_D)}
\end{equation}

\noindent is the renormalization due to the presence of the repulsive interaction. The effect of the repulsive Coulomb potential is therefore to renormalize the strength \(\lambda\) of the attractive potential in the critical temperature formula. At sufficiently large renormalization (\(\mu^* \geq \lambda\)), the analysis breaks down, and there is no superconducting instability. 

After solving the gap equation with different Coulomb repulsion strengths, we fit the critical temperature to a function in the form

\begin{equation}
k_B T_c = 1.14 \; \hbar \omega_D \exp \left\{ -\frac{1}{\lambda - \frac{au}{1+abu}} \right\},
\end{equation}

\noindent with two fitting parameters \(a\) and \(b\) in addition to the electron-phonon coupling strength \(\lambda\), which is fixed by the critical temperature at zero repulsive Coulomb interaction.

\bibliography{ref}
\end{document}